\documentclass[prl,preprint,superscriptaddress]{revtex4-2}
\usepackage[dvipdfmx]{graphicx}
\usepackage{dcolumn}
\usepackage{bm}
\usepackage{braket}

\usepackage{color}

\begin{document}

\title{Fluctuated spin--orbital texture of Rashba-split surface states in real and reciprocal space}

\author{Takuto Nakamura}
\email{t.nakamura@fbs.osaka-u.ac.jp}
\affiliation{Graduate School of Frontier Biosciences, Osaka University, Suita 565-0871, Japan}
\affiliation{Department of Physics, Graduate School of Science, Osaka University, Toyonaka 560-0043, Japan}
\author{Yoshiyuki Ohtsubo}
\email{y\_oh@qst.go.jp}
\affiliation{Institute for Advances Synchrotron Light Source, National Institutes for Quantum Science and Technology, Sendai 980-8579, Japan}
\affiliation{Graduate School of Frontier Biosciences, Osaka University, Suita 565-0871, Japan}
\affiliation{Department of Physics, Graduate School of Science, Osaka University, Toyonaka 560-0043, Japan}
\author{Ayumi Harasawa}
\affiliation{Institute for Solid State Physics, The University of Tokyo, Kashiwa 277-8581, Japan}
\author{Koichiro Yaji}
\affiliation{National Institute for Materials Science (NIMS), Tsukuba, Ibaraki 305-0003, Japan}
\author{Shik Shin}
\author{Fumio Komori}
\affiliation{Institute for Solid State Physics, The University of Tokyo, Kashiwa 277-8581, Japan}
\author{Shin-ichi Kimura}
\email{kimura@fbs.osaka-u.ac.jp}
\affiliation{Graduate School of Frontier Biosciences, Osaka University, Suita 565-0871, Japan}
\affiliation{Department of Physics, Graduate School of Science, Osaka University, Toyonaka 560-0043, Japan}
\affiliation{Institute for Molecular Science, Okazaki 444-8585, Japan}

\date{\today}

\begin{abstract}
Spin-orbit interaction (SOI) in low-dimensional systems, namely Rashba systems and the edge states of topological materials, is extensively studied in this decade as a promising source to realize various fascinating spintronic phenomena, such as the source of the spin current and spin-mediated energy conversion.
Here, we show the odd fluctuation in the spin--orbital texture in a surface Rashba system on Bi/InAs(110)-(2$\times$1) by spin- and angle-resolved photoelectron spectroscopy and a numerical simulation based on a density-functional theory (DFT) calculation.
The surface state shows a paired parabolic dispersion with the spin degeneracy lifted by the Rashba effect.
Although its spin polarization should be fixed in a particular direction based on the Rashba model, the observed spin polarization varies greatly and even reverses its sign depending on the wavenumber.
DFT calculations also reveal that the spin directions of two inequivalent Bi chains on the surface change from nearly parallel (canted-parallel) to anti-parallel in real space in the corresponding wavevector region.
These results point out an oversimplification of the nature of spin in Rashba and Dirac systems and provide more freedom than expected for spin manipulation of photoelectrons.
\end{abstract}


\maketitle
The symmetry breaking of a system often causes fertile unconventional physical phenomena, such as nematic superconductivity appearing with rotational asymmetry \cite{Kasahara12} and orbital-angular-momentum (OAM) polarization of graphene Dirac-cone states caused by the asymmetry between sublattices of honeycomb lattice (pseudospin) \cite{Gierz12}.
In particular, spin--orbit interaction (SOI) in two-dimensional (2D) surface systems, namely, Rashba effect \cite{Rashba84} and edge states of topological materials \cite{Kane10}, is a promising source to host, create, or convert spin-polarized electrons without magnetic materials and fields \cite{Manchon15, Han18}.

A schematic of SOI-caused helical spin polarization in a 2D system is depicted in Fig. 1(a), where surface electrons are polarized toward the orientation perpendicular to both the wavevector and the surface normal.
In this model, the potential gradient at the surface is assumed to be parallel to the surface normal.
An electron moving in such a potential gradient experiences an effective magnetic field $\vec{B}_{eff} \propto \vec{E}\times\vec{k}$ resulting in Zeeman-like splitting and spin polarizations parallel and anti-parallel to $\vec{B}_{eff}$ \cite{Rashba84}.
This can be regarded as the outcome of the symmetry breaking in the 2D system along the out-of-plane orientation.
The in-plane asymmetry also contributes to the surface spin polarization, although it is not included in the simple Rashba model considered above.
For example, out-of-plane spin polarization in triangular surface lattices, \textit{i.e.}, valley polarization, is caused by the lack of in-plane two-fold rotation symmetry \cite{Sakamoto09, Souma11, Suzuki14}.

In the simple Rashba model, one spin-split branch of the surface band always holds one spin orientation: clockwise (CW) or counter CW (CCW) in the reciprocal space.
This assumption was consistent with various SARPES data reported by early 2010s except for few cases with valley polarization \cite{Sakamoto09, Okuda13} and is widely used in the interpretation of surface-spin-mediated energy conversion experiments \cite{Han18}.
However, recent studies, by using the combination of the spin- and angle-resolved photoelectron spectroscopy (SARPES) and density-functional theory (DFT) calculation, revealed the contribution of complex combinations of spin and orbital terms in the wavefunctions to 2D spin-polarized surface states, even in a single surface band \cite{Zhu14, Kuroda16, Yaji17}.
These studies were mainly performed by using high-symmetry surface systems, such as $C_{3v}$ symmetry with three-fold rotation and three mirror planes, implying that more variants of spin and orbital textures would appear in lower-symmetry systems.
Furthermore, the minimal features of Rashba-type SOI in the surface electronic structure should be distinguished from coincidental coexisting features realized by surface symmetry operations, which do not always exist in general surface systems.

In this work, we studied the spin and orbital texture in the surface electronic states of Bi/InAs(110)-(2$\times$1) by SARPES and a numerical simulation based on DFT calculations.
The (2$\times$1) superlattice of Bi on InAs(110) consists of two non-equivalent and tilted Bi chains, as depicted in Fig. 1(b) \cite{Betti99}.
This surface structure holds a small number of surface symmetry operations, only one mirror plane, and no rotation, and thus, is one of the ideal test case to study the minimal feature of surface SOI.
Actually, the surface state exhibited an unconventional spin fluctuation for the first time.
The sign on the surface spin polarization modulated independently in the two nonequivalent surface atomic chains and even in a single band, its sign inverted depending on the wavenumber.
These results point out an oversimplification of the nature of spin in surface 2D systems, and suggest more freedom than expected for spin manipulation of photoelectrons.

\section*{Results}
\subsection*{Photoelectron spectra of spin-polarized surface electronic states}
Figures 2(a) and 2(b) show conventional (spin-integrated) ARPES band dispersions along $\bar{\Gamma}$ -- $\bar{\rm X}$ using \textit{s}- and \textit{p}-polarized photons, respectively.
The experimental geometry and definitions of the incident-photon polarizations of the (S)ARPES experiment are shown in Fig. 2(e).
As shown in a previous work \cite{Nakamura18}, surface bands (S and S') were observed below the Fermi level ($E_{\rm F}$) with paired parabolic dispersion around the surface Brillouin zone center ($k_{y//[\bar{1}10]}$ = 0 \AA$^{-1}$).
The photoelectron intensity of S with \textit{s}-polarization was diminished around $k_{y//[\bar{1}10]}$ = 0 \AA$^{-1}$, suggesting an influence of the photoexcitation selection rule \cite{PEStext}.
Assuming the nearly-free-electron final state, photoelectrons from the initial states with odd and even parities with respect to the measurement plane (see Fig. 2(e)) should be excited by \textit{s}- and \textit{p}-polarized photons, respectively.
This suggests that the inner part of the parabolic S band is mainly composed of wavefunctions with even symmetry, whereas the outer part is from multiple wavefunctions with both even and odd symmetries.

Figures 2(c1--c3) and 2(d1--d3) show polarization-dependent SARPES map obtained by using \textit{s}- and \textit{p}-polarized photons, respectively.
These are two-dimensional color plots of photoelectron intensity and magnitude of spin polarization (see Figs. 2(f1--f3) for their definitions).
The corresponding spin-resolved energy distribution curves are also shown in supplementary Fig. S2 \cite{SM}.
The band dispersion obtained by SARPES intensities shown in Figs. 2(c1--d3) agrees well with the spin-integrated ones in Figs. 2(a, b).
Figures 2(c1, d1) show the spin polarization along the $S_x$ direction.
This orientation is in-plane and perpendicular to the wavevector, which is expected from the simple 2D model as depicted in Fig. 1(a).
However, in contrast to the simple model, Fig. 2(d1) shows that the $S_x$ spin polarization direction varies in a single parabolic band (S) with the same sign of $k_{y//[\bar{1}10]}$.
Such a peculiar spin modulation is not observed in Fig. 2(c1), indicating the effect of incident photon polarizations  (\textit{s} and \textit{p}).
Thus far, it has been reported in some topological materials that to switch the photon polarization could change the spin polarization of photoelectrons \cite{Zhu14, Yaji17}.
However, the spin modulation in a single band with the same polarization, which is observed herein with \textit{p} polarization, has not been reported yet, to the best of our knowledge.
In contrast to S, another spin-polarized band S' shows uni-directional polarization in the entire $k_{y//[\bar{1}10]}$ range.
Therefore, we focus on the spin polarization of S in the following part, in order to investigate this peculiar behavior.

In addition to $S_x$, spin polarizations of the other spin directions $S_y$ and $S_z$ were observed in Figs. 2(c2--c3) and 2(d2--d3).
$S_z$ is almost zero with \textit{s}-polarized photons but has a finite value with \textit{p}-polarized photons. 
Although the $S_z$ polarization normal to the surface is not expected from the simplest model, as shown in Fig. 1(a), this polarization is allowed with an in-plane potential gradient, as reported in some 2D systems without two-fold rotation \cite{Sakamoto09, Souma11, Suzuki14}.
However, the peculiar spin reversal as that of $S_x$ in the single band (Fig. 2(d3)) has not been reported so far.
Moreover, in this measurement, an evident spin polarization parallel to the wavevector ($S_y$) is also observed with both photon polarizations, as shown in Figs. 2(c2, d2).
The spin polarization along this orientation is not expected from the Rashba model because the effective magnetic field originates from the cross product $ \vec{E}\times\vec{k}$, which is always orthogonal to $\vec{k}$.
The Bi/InAs(110)-(2$\times$1) surface does not contain any magnetic element, and hence, the observed complex spin polarization cannot be explained solely from the initial states.
Thus, it is suggested that the the photo-excitation process is playing some role for the observed complex spin texture.

\subsection*{Spin polarizations of the ground states obtained from theoretical calculations}
To reveal the initial-state spin texture, we calculated the surface band structures by DFT, as shown in Fig. 3.
The sizes of the circles in Figs. 3(a) and 3(b) indicate the contribution of surface Bi atoms.
Color scales (defined in Fig. 3(e)) show the degree of the spin polarizations along $S_x$ and $S_z$.
The shapes of the surface bands (S and S') are qualitatively consistent with the results observed by ARPES and SARPES.
All the spin-polarized structures are reversed with the sign of the wavenumber, indicating that the electronic structure obeys time-reversal symmetry.
It should be noted that $S_y$ spin polarization was not obtained in the DFT calculation, as expected from the time-reversal symmetry of the system confirmed above. 

To obtain further insight into the surface spin texture, we decomposed the contributions to the surface bands from two non-equivalent surface Bi chains: Bi1-2 and Bi3-4 (see Fig. 1(b) for the definition).
The radii of the circles in Figs. 3(c1--d2) show that both Bi chains contribute to surface band S with nearly the same size.
The in-plane spin polarization $S_x$ for chains Bi1-2 and Bi3-4 are shown in Figs. 3(c1) and 3(c2), respectively.
Although they point [00$\bar{1}$] in most of the region of positive $k_y$, such spin polarization decreases near $\bar{\Gamma}$ and even changes the direction (pointing [001]) for Bi3-4.
On the contrary, the out-of-plane spin polarizations $S_z$ of Bi chains are opposite to each other, as shown in Figs. 3(d1--d2).
This indicates that each Bi chain has independent spin polarization, even at the same (\textit{E}, $\vec{k}$) points.
In contrast to $S_x$, the $S_z$ spin polarization is large around $\bar{\Gamma}$ and decreases in $k_y >$ 0.2 \AA$^{-1}$.
This behavior of $S_x$ and $S_z$ indicates the rotation of the spin polarization of S depending on $k_y$.

To visualize the spin texture in real space, the spin rotation angle depending on the wavenumber was calculated for the S band.
The results are shown in Fig. 3(f).
The spin rotation angle is defined as $S_z$ = +1 to be the origin (0$^\circ$) and $S_x$ = +1 to be 90$^\circ$.
Interestingly, the spin rotation angle of the S band varies continuously in the xz-plane depending on the wavenumber, and the sign of the rotation is opposite between Bi1-2 and Bi3-4 in the region from the $\bar{\Gamma}$ point ($k_{y//[\bar{1}10]}$ = 0.0 \AA$^{-1}$) to near the top of the parabola ($k_{y//[\bar{1}10]} \sim$ 0.2 \AA$^{-1}$).
This indicates that the sign reversal of the $S_x$ polarization in Bi3-4 is the outcome of such a continuous spin rotation in the xz-plane.
Figure 3(h) depicts the spin orientation of each Bi chain in real space at some specific wavenumbers indicated in Fig. 3(f).
Near $\bar{\Gamma}$ (area (I)), the spin orientation in each chain is completely different.
In area (II), slightly closer to the apex of the parabolic band, the polarization direction is anti-parallel, showing anti-ferromagnetic spin order.
Near the apex of the parabola (area III), the spins of the two Bi chains become nearly parallel with a small $S_z$ component.
In this $k_y$ region, the surface spin texture is similar to the spin structure expected for a normal Rashba-type spin texture, namely the ‘‘ferromagnetic’’ case, as compared to region (II).
Wavenumber-dependent spin rotation in the Rashba system has already been reported \cite{Sakamoto09}, but such fertile spin orders, from the anti-ferromagnetic to nearly ferromagnetic cases between two non-equivalent atomic chains, are not expected from ordinary Rashba systems.

\subsection*{Comparison between the observed photoelectron spectra and the calculated ground states}
Here, we compare the calculated spin texture of the initial surface states with the experimentally observed spin-polarized photoelectron spectra.
For this purpose, we decomposed the calculated Bi orbital wavefunctions to those odd ($p_x$) and even ($p_y$ and $p_z$) with respect to the measurement plane, as shown in Figs. 4(a1--b2).
The odd and even initial states should correspond to the photoelectrons excited by \textit{s}- and \textit{p}-polarized photons shown in Fig. 2, respectively, based on the selection rule of the electric dipole photo-excitation.
The radii of the circles in Figs. 4(a1--b2) are proportional to the surface Bi wavefunctions with odd or even parities.
Odd (even) Bi wavefunctions shown in Figs. 4(a1, b1) (4(a2, b2)) have a large contribution around (away from) $\bar{\Gamma}$ for surface band S.
This trend is consistent with the small photoelectron intensity around $\bar{\Gamma}$ by \textit{s}-polarized photons (Fig. 2(a)).
As for the spin polarization, the in-plane orientation ($S_x$) of S could reproduce the observed results.
Although the spins of the odd wavefunctions (Fig. 4(a1)) are intense and monotonic, those of the even wavefunctions (Fig. 4(a2)) show the sign inversion from 0 \AA$^{-1}$ ($\bar{\Gamma}$) to 0.2 \AA$^{-1}$ (near the top of the parabolic band) in surface band S.
In contrast to the in-plane case, the situation of the out-of-plane spins ($S_z$, shown in Figs. 4(d1, d2)) is complicated.
Although nearly negligible values for the odd wavefunctions are consistent with the observed results (Figs. 2(c3)), the observed inversion (Figs. 2(d3)) is not reproduced for the even wavefunctions.
Together with the exactly zero polarization parallel to $k_y$ ($S_y$) in the calculation, it is obvious that the initial state calculation is insufficient for understanding the observed spin polarization of photoelectrons.

The photoelectron interference effect in the spin-polarized states has been already discussed for topological materials Bi and Bi$_2$Se$_3$ \cite{Zhu13, Zhu14, Kuroda16, Yaji17}.
If two or more wavefunctions with different spin polarizations are photo-excited simultaneously, the interference between them can result in the unique spin polarization of final-state photoelectrons.
As shown in Fig. 3(f), Bi/InAs(110)-(2$\times$1) has two non-equivalent Bi chains with different spin orientations at each $k_y$.
We adopted these two Bi chains as the different initial states and calculated their interference numerically during the spin-dependent photo-excitation process in a similar manner to earlier works \cite{Zhu13, Yaji17} (see supplementary Note 1 for details).
Figures 5(a1) and 5(a2) are the spin polarization of the initial states of the S bands obtained from our DFT calculation with the odd and even symmetries, respectively.
Figure 5(b1) (5(b2)) depicts the simulated photoelectron spin polarizations corresponding to the initial states shown in Fig. 5(a1) (5(a2)).
Figures 5(d1) and 5(d2) are the corresponding spin polarizations of the observed photoelectrons. 
For the odd states (Figs. 5(a1) and 5(d1)), monotonic polarizations with sizable $S_x$ and small $S_z$ of the calculation agree well with the experimental results.
A decrease in $S_x$ in the vicinity of $\bar{\Gamma}$ that appears in the experimental result does not appear in the calculation.
The possible reason is that the experimental value is not so reliable owing to the vanishing photoelectron intensities.
For the even states (Figs. 5(a2) and 5(d2)), the calculated directions of $S_x$ and $S_z$ are consistent with the corresponding experimental ones.
The inversion of $S_z$ is absent in the initial states (Fig. 5(a2)) but is propagated from $S_x$ by the photoelectron interference.
It should be noted that the spin polarization parallel to \textit{k} ($S_y$) is also obtained in the calculation, in contrast to its absence in the initial states.
Although this calculation is insufficient for reproducing some quantitative values, such as the spin polarization ratio $S_x$/$S_z$ for the even states, it is sufficient for capturing the qualitative behavior of this complicated spin texture.

\section*{Discussion}
The consistency of the numerical calculation and SARPES results supports the peculiar spin orders between neighboring Bi chains, fluctuating depending on the wavenumber, as shown in Fig. 3.
It should be noted that a new mechanism is not required to explain such a spin texture in the surface ground states.
As already observed in Weyl semimetals \cite{Xu16,Feng16}, spin polarization derived from SOI can point arbitrary orientations as far as the surface symmetry operations permit.
The reason why the current result seems unnatural could be because most of earlier works on the surface Rashba states were performed in simple surfaces.
Therein, the number of non-equivalent atoms was smaller, and the number of symmetry operations was larger than in the present Bi/InAs(110)-($2\times1$) case with only one mirror plane. 
The current results also exhibit that the excited electrons could have various, almost arbitrary spin polarizations depending on the photo-excitation conditions.
Not only the fine tuning of photon polarization \cite{Yaji17} but also the wavenumber selection with the electronic states derived from many non-equivalent wavefunctions can provide highly spin polarized electrons with the polarization orientations on demand.

Surface band S of Bi/InAs(110)-(2$\times$1) appears to obey surface Rashba effect, a paired parabolic dispersion with helical spin polarization \cite{Nakamura18}.
However, the current result revealed that the actual spin and orbital constitution is considerably more complex than the simple Rashba model, as shown in Fig. 1.
This indicates that such a model is too simple to represent the realistic system.
Actually, other typical Rashba and topological systems, such as Bi$_2$Se$_3$ \cite{Zhu13, Kuroda16}, Bi(111) \cite{Yaji17}, and Bi/Ag(111) \cite{Noguchi17}, have multiple spin and orbital components.
Because SOI itself is an interaction to mix the off-diagonal orbital characters with different spins \cite{Chadi77, Petersen00}, the complex spin--orbital entanglement would be the general characteristics of the low-dimensional electronic states influenced by SOI.
Some theoretical studies claim that such a mixture is essential for Rashba-type SOI \cite{Premper07, Ishida14}.

\section*{Methods }

\subsection*{Sample preparation}
The surface of InAs(110) substrates was cleaned by repeated cycles of Ar ion sputtering (1 keV) and annealing (700 K).
After the cleaning, a sharp (1$\times$1) electron diffraction pattern was confirmed by using the low-energy electron diffraction  (LEED).
Next, a few monolayers of Bi were deposited on InAs(110)-(1$\times$1) from a Knudsen cell at room temperature.
Subsequent annealing at 600 K for 20 min resulted in a Bi/InAs(110) surface exhibiting a sharp (2$\times$1) surface reconstruction pattern.
The LEED diffraction patterns are shown in supplementary Fig. S1 \cite{SM}.

\subsection*{Laser-ARPES and SARPES experimental setup}
ARPES and SARPES measurements were performed at the Institute for Solid State Physics, the University of Tokyo with a linearly polarized laser source ($h\nu$ = 6.994 eV) \cite {Yaji16}.
The photoelectrons were detected along the $\bar{\Gamma}$ -- $\bar{\rm X}$ high-symmetry line in the (2$\times$1) surface Brillouin zone, as shown in Fig. 2(e).
In this experimental geometry, the planes of photon incidence and photoelectron detection were common.
The energy resolution and the position of $E_{\rm F}$ were calibrated by the Fermi edge of a Cu block attached to the samples.
The energy resolution for the ARPES (SARPES) measurements was set to $\sim$9 meV ($\sim$20 meV).
The effective Sherman function of a very-low-energy-electron-diffraction (VLEED)-type spin detector was set to 0.27.
The sample temperature was kept at 45 K during the ARPES and SARPES experiments.

\subsection*{Calculation of photoelectron spin polarization}
The initial states and their spin polarization of the Bi/InAs(110) surface were obtained by using a DFT calculation based on the `` augmented plane wave $+$ local orbitals '' method implemented in the WIEN2k code with SOI considered \cite{WIEN2k}.
We adopted the modified Becke--Johnson potential combined with a local density approximation to construct the exchange and correlation potentials \cite{mBJ1, mBJ2}.
The surface atomic structure was modeled by 20 layers of InAs with a surface covered with the (2$\times$1) zig-zag Bi chains.
The surface atomic structure was energetically optimized down to the third In and As layers, and the rest of the substrate atom positions were fixed to those in the bulk InAs single crystal.
The eventual surface atomic structure model was almost identical to the model obtained experimentally \cite{Betti99}.

The spin polarization of the photoelectrons was numerically calculated considering the interference in the final-state spinors.
For this calculation, we assumed the photoexcitation process from the non-equivalent initial-state wavefunctions, the two different Bi chains, obtained by the DFT calculation above.
The final state was set as a spin-integrated free electron.
The spinor interference was calculated in a similar manner to refs. \cite{Zhu14, Yaji17} (see supplementary note 1 \cite{SM} for details).

\section*{Acknowledgements}
The SARPES measurements were jointly conducted with ISSP, the University of Tokyo.
This work was also supported by JSPS KAKENHI (Grants Nos. JP20K03859, JP19H01830, JP18K03484, and JP20H04453).

\section*{Author contributions}
T.N. and Y.O. conducted the SARPES experiments with assistance from A.H., K.Y., S.S., and F.K.
Y.O. performed the DFT calculation, and its numerical analysis to simulate photoelectron spin polarizations was performed by T.N. and Y.O.
T.N., Y.O., and S.-i.K. wrote the text and were responsible for the overall direction of the research project.
All authors contributed to the scientific planning and discussions.

\section*{Competing financial interests}
The authors declare no competing financial interests.

\section*{Data availability}
The dataset generated during and/or analyzed during the current study are available from the corresponding author on reasonable request.

\bibliographystyle{naturemag}
\bibliography{bibtest}

\providecommand{\noopsort}[1]{}\providecommand{\singleletter}[1]{#1}%
\begin{thebibliography}{10}
\expandafter\ifx\csname url\endcsname\relax
  \def\url#1{\texttt{#1}}\fi
\expandafter\ifx\csname urlprefix\endcsname\relax\def\urlprefix{URL }\fi
\providecommand{\bibinfo}[2]{#2}
\providecommand{\eprint}[2][]{\url{#2}}

\bibitem{Kasahara12}
\bibinfo{author}{Kasahara, S.} \emph{et~al.}
\newblock \bibinfo{title}{{Electronic nematicity above the structural and
  superconducting transition in BaFe$_2$(As$_{1-x}$P$_x$)$_2$}}.
\newblock \emph{\bibinfo{journal}{Nature}} \textbf{\bibinfo{volume}{486}},
  \bibinfo{pages}{382--385} (\bibinfo{year}{2012}).

\bibitem{Gierz12}
\bibinfo{author}{Gierz, I.}, \bibinfo{author}{Lindroos, M.},
  \bibinfo{author}{H{\"o}chst, H.}, \bibinfo{author}{Ast, C.~R.} \&
  \bibinfo{author}{Kern, K.}
\newblock \bibinfo{title}{{Graphene sublattice symmetry and isospin determined
  by circular dichroism in angle-resolved photoemission spectroscopy}}.
\newblock \emph{\bibinfo{journal}{Nano letters}} \textbf{\bibinfo{volume}{12}},
  \bibinfo{pages}{3900--3904} (\bibinfo{year}{2012}).

\bibitem{Rashba84}
\bibinfo{author}{Bychkov, Y.~A.} \& \bibinfo{author}{Rashba, E.~I.}
\newblock \bibinfo{title}{{Oscillatory effects and the magnetic susceptibility
  of carriers in inversion layers}}.
\newblock \emph{\bibinfo{journal}{Journal of physics C: Solid state physics}}
  \textbf{\bibinfo{volume}{17}}, \bibinfo{pages}{6039} (\bibinfo{year}{1984}).

\bibitem{Kane10}
\bibinfo{author}{Hasan, M.~Z.} \& \bibinfo{author}{Kane, C.~L.}
\newblock \bibinfo{title}{{Colloquium: Topological insulators}}.
\newblock \emph{\bibinfo{journal}{Rev. Mod. Phys.}}
  \textbf{\bibinfo{volume}{82}}, \bibinfo{pages}{3045--3067}
  (\bibinfo{year}{2010}).

\bibitem{Manchon15}
\bibinfo{author}{Manchon, A.}, \bibinfo{author}{Koo, H.~C.},
  \bibinfo{author}{Nitta, J.}, \bibinfo{author}{Frolov, S.~M.} \&
  \bibinfo{author}{Duine, R.~a.}
\newblock \bibinfo{title}{{New perspectives for Rashba spin–orbit coupling}}.
\newblock \emph{\bibinfo{journal}{Nature Materials}}
  \textbf{\bibinfo{volume}{14}}, \bibinfo{pages}{871--882}
  (\bibinfo{year}{2015}).

\bibitem{Han18}
\bibinfo{author}{Han, W.}, \bibinfo{author}{Otani, Y.} \&
  \bibinfo{author}{Maekawa, S.}
\newblock \bibinfo{title}{{Quantum materials for spin and charge conversion}}.
\newblock \emph{\bibinfo{journal}{npj Quantum Materials}}
  \textbf{\bibinfo{volume}{3}}, \bibinfo{pages}{1--16} (\bibinfo{year}{2018}).

\bibitem{Sakamoto09}
\bibinfo{author}{Sakamoto, K.} \emph{et~al.}
\newblock \bibinfo{title}{{Abrupt Rotation of the Rashba Spin to the Direction
  Perpendicular to the Surface}}.
\newblock \emph{\bibinfo{journal}{Phys. Rev. Lett.}}
  \textbf{\bibinfo{volume}{102}}, \bibinfo{pages}{096805}
  (\bibinfo{year}{2009}).

\bibitem{Souma11}
\bibinfo{author}{Souma, S.} \emph{et~al.}
\newblock \bibinfo{title}{{Direct Measurement of the Out-of-Plane Spin Texture
  in the Dirac-Cone Surface State of a Topological Insulator}}.
\newblock \emph{\bibinfo{journal}{Phys. Rev. Lett.}}
  \textbf{\bibinfo{volume}{106}}, \bibinfo{pages}{216803}
  (\bibinfo{year}{2011}).

\bibitem{Suzuki14}
\bibinfo{author}{Suzuki, R.} \emph{et~al.}
\newblock \bibinfo{title}{{Valley-dependent spin polarization in bulk MoS$_2$
  with broken inversion symmetry}}.
\newblock \emph{\bibinfo{journal}{Nature nanotechnology}}
  \textbf{\bibinfo{volume}{9}}, \bibinfo{pages}{611--617}
  (\bibinfo{year}{2014}).

\bibitem{Okuda13}
\bibinfo{author}{Okuda, T.} \& \bibinfo{author}{Kimura, A.}
\newblock \bibinfo{title}{{Spin- and Angle-Resolved Photoemission of Strongly
  Spin–Orbit Coupled Systems}}.
\newblock \emph{\bibinfo{journal}{Journal of the Physical Society of Japan}}
  \textbf{\bibinfo{volume}{82}}, \bibinfo{pages}{021002}
  (\bibinfo{year}{2013}).

\bibitem{Zhu14}
\bibinfo{author}{Zhu, Z.-H.} \emph{et~al.}
\newblock \bibinfo{title}{{Photoelectron Spin-Polarization Control in the
  Topological Insulator ${\mathrm{Bi}}_{2}{\mathrm{Se}}_{3}$}}.
\newblock \emph{\bibinfo{journal}{Phys. Rev. Lett.}}
  \textbf{\bibinfo{volume}{112}}, \bibinfo{pages}{076802}
  (\bibinfo{year}{2014}).

\bibitem{Kuroda16}
\bibinfo{author}{Kuroda, K.} \emph{et~al.}
\newblock \bibinfo{title}{{Coherent control over three-dimensional spin
  polarization for the spin-orbit coupled surface state of
  ${\mathrm{Bi}}_{2}{\mathrm{Se}}_{3}$}}.
\newblock \emph{\bibinfo{journal}{Phys. Rev. B}} \textbf{\bibinfo{volume}{94}},
  \bibinfo{pages}{165162} (\bibinfo{year}{2016}).

\bibitem{Yaji17}
\bibinfo{author}{Yaji, K.} \emph{et~al.}
\newblock \bibinfo{title}{{Spin-dependent quantum interference in photoemission
  process from spin-orbit coupled states}}.
\newblock \emph{\bibinfo{journal}{Nature communications}}
  \textbf{\bibinfo{volume}{8}}, \bibinfo{pages}{1--6} (\bibinfo{year}{2017}).

\bibitem{Betti99}
\bibinfo{author}{Betti, M.~G.} \emph{et~al.}
\newblock \bibinfo{title}{{$(1\ifmmode\times\else\texttimes\fi{}2)$ Bi chain
  reconstruction on the InAs(110) surface}}.
\newblock \emph{\bibinfo{journal}{Phys. Rev. B}} \textbf{\bibinfo{volume}{59}},
  \bibinfo{pages}{15760--15765} (\bibinfo{year}{1999}).

\bibitem{Nakamura18}
\bibinfo{author}{Nakamura, T.} \emph{et~al.}
\newblock \bibinfo{title}{{Giant Rashba splitting of quasi-one-dimensional
  surface states on Bi/InAs(110)-$(2\ifmmode\times\else\texttimes\fi{}1)$}}.
\newblock \emph{\bibinfo{journal}{Phys. Rev. B}} \textbf{\bibinfo{volume}{98}},
  \bibinfo{pages}{075431} (\bibinfo{year}{2018}).

\bibitem{PEStext}
\bibinfo{author}{H\"{u}fner, S.}
\newblock \emph{\bibinfo{title}{{Photoelectron Spectroscopy}}}
  (\bibinfo{publisher}{Springer-Verlag Berlin Heidelberg},
  \bibinfo{year}{2003}).

\bibitem{SM}
\bibinfo{note}{Supplementary Materials contain the additional experimental
  dataset, LEED patterns and SARPES spectra, and detailed calculation method to
  simulate the photoelectron spin polarization.}

\bibitem{Zhu13}
\bibinfo{author}{Zhu, Z.-H.} \emph{et~al.}
\newblock \bibinfo{title}{{Layer-By-Layer Entangled Spin-Orbital Texture of the
  Topological Surface State in ${\mathrm{Bi}}_{2}{\mathrm{Se}}_{3}$}}.
\newblock \emph{\bibinfo{journal}{Phys. Rev. Lett.}}
  \textbf{\bibinfo{volume}{110}}, \bibinfo{pages}{216401}
  (\bibinfo{year}{2013}).

\bibitem{Xu16}
\bibinfo{author}{Xu, S.-Y.} \emph{et~al.}
\newblock \bibinfo{title}{{Spin Polarization and Texture of the Fermi Arcs in
  the Weyl Fermion Semimetal TaAs}}.
\newblock \emph{\bibinfo{journal}{Phys. Rev. Lett.}}
  \textbf{\bibinfo{volume}{116}}, \bibinfo{pages}{096801}
  (\bibinfo{year}{2016}).

\bibitem{Feng16}
\bibinfo{author}{Feng, B.} \emph{et~al.}
\newblock \bibinfo{title}{{Spin texture in type-II Weyl semimetal
  ${\mathrm{WTe}}_{2}$}}.
\newblock \emph{\bibinfo{journal}{Phys. Rev. B}} \textbf{\bibinfo{volume}{94}},
  \bibinfo{pages}{195134} (\bibinfo{year}{2016}).

\bibitem{Noguchi17}
\bibinfo{author}{Noguchi, R.} \emph{et~al.}
\newblock \bibinfo{title}{{Direct mapping of spin and orbital entangled wave
  functions under interband spin-orbit coupling of giant Rashba spin-split
  surface states}}.
\newblock \emph{\bibinfo{journal}{Phys. Rev. B}} \textbf{\bibinfo{volume}{95}},
  \bibinfo{pages}{041111} (\bibinfo{year}{2017}).

\bibitem{Chadi77}
\bibinfo{author}{Chadi, D.~J.}
\newblock \bibinfo{title}{{Spin-orbit splitting in crystalline and
  compositionally disordered semiconductors}}.
\newblock \emph{\bibinfo{journal}{Phys. Rev. B}} \textbf{\bibinfo{volume}{16}},
  \bibinfo{pages}{790--796} (\bibinfo{year}{1977}).

\bibitem{Petersen00}
\bibinfo{author}{Petersen, L.} \& \bibinfo{author}{Hedeg{\aa}rd, P.}
\newblock \bibinfo{title}{{A simple tight-binding model of spin–orbit
  splitting of sp-derived surface states}}.
\newblock \emph{\bibinfo{journal}{Surface Science}}
  \textbf{\bibinfo{volume}{459}}, \bibinfo{pages}{49--56}
  (\bibinfo{year}{2000}).

\bibitem{Premper07}
\bibinfo{author}{Premper, J.}, \bibinfo{author}{Trautmann, M.},
  \bibinfo{author}{Henk, J.} \& \bibinfo{author}{Bruno, P.}
\newblock \bibinfo{title}{{Spin-orbit splitting in an anisotropic
  two-dimensional electron gas}}.
\newblock \emph{\bibinfo{journal}{Phys. Rev. B}} \textbf{\bibinfo{volume}{76}},
  \bibinfo{pages}{073310} (\bibinfo{year}{2007}).

\bibitem{Ishida14}
\bibinfo{author}{Ishida, H.}
\newblock \bibinfo{title}{{Rashba spin splitting of Shockley surface states on
  semi-infinite crystals}}.
\newblock \emph{\bibinfo{journal}{Phys. Rev. B}} \textbf{\bibinfo{volume}{90}},
  \bibinfo{pages}{235422} (\bibinfo{year}{2014}).

\bibitem{Yaji16}
\bibinfo{author}{Yaji, K.} \emph{et~al.}
\newblock \bibinfo{title}{{High-resolution three-dimensional spin- and
  angle-resolved photoelectron spectrometer using vacuum ultraviolet laser
  light}}.
\newblock \emph{\bibinfo{journal}{Review of Scientific Instruments}}
  \textbf{\bibinfo{volume}{87}}, \bibinfo{pages}{053111}
  (\bibinfo{year}{2016}).

\bibitem{WIEN2k}
\bibinfo{author}{Blaha, P.} \emph{et~al.}
\newblock \bibinfo{title}{{WIEN2k: An APW+lo program for calculating the
  properties of solids}}.
\newblock \emph{\bibinfo{journal}{The Journal of Chemical Physics}}
  \textbf{\bibinfo{volume}{152}}, \bibinfo{pages}{074101}
  (\bibinfo{year}{2020}).

\bibitem{mBJ1}
\bibinfo{author}{Becke, A.~D.} \& \bibinfo{author}{Johnson, E.~R.}
\newblock \bibinfo{title}{{A simple effective potential for exchange}}.
\newblock \emph{\bibinfo{journal}{The Journal of Chemical Physics}}
  \textbf{\bibinfo{volume}{124}}, \bibinfo{pages}{221101}
  (\bibinfo{year}{2006}).

\bibitem{mBJ2}
\bibinfo{author}{Tran, F.} \& \bibinfo{author}{Blaha, P.}
\newblock \bibinfo{title}{Accurate band gaps of semiconductors and insulators
  with a semilocal exchange-correlation potential}.
\newblock \emph{\bibinfo{journal}{Phys. Rev. Lett.}}
  \textbf{\bibinfo{volume}{102}}, \bibinfo{pages}{226401}
  (\bibinfo{year}{2009}).

\end{thebibliography}


\newpage

\begin{figure}[p]
\includegraphics[width=80mm]{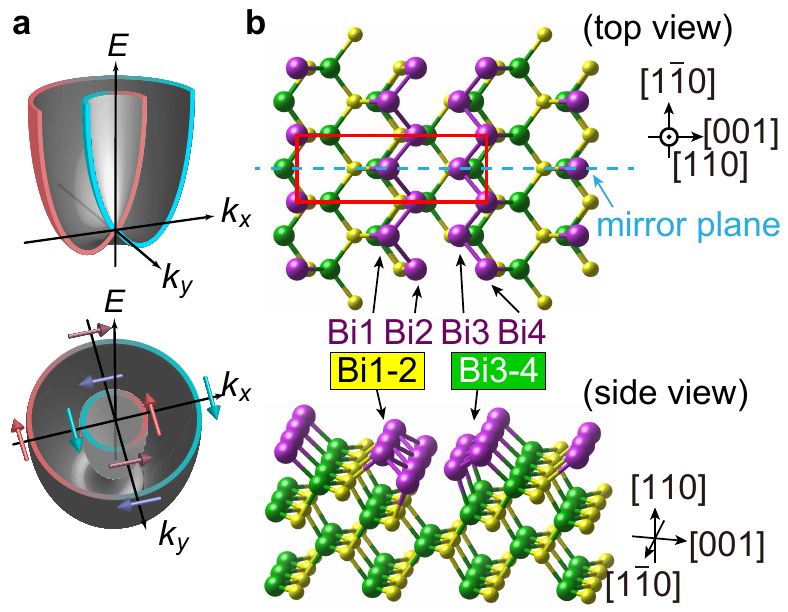}
\caption{\label{figure 1}
{\bf Schematic of conventional Rashba-type spin-split bands and surface atomic structure of Bi/InAs(110)-(2$\times$1).}
(a) Ideal two-dimensional spin-split electronic bands due to Rashba-type SOI. 
The spin polarization direction is perpendicular to the wavevector $\vec{k}$, exhibiting a helical spin polarization in reciprocal space.
(b) Surface atomic structure of Bi/InAs(110)-(2$\times$1).
Tilted atomic chains of Bi on the InAs(110) substrate form the (2$\times$1) surface superlattice, which consists of four non-equivalent Bi atoms, Bi1, Bi2, Bi3, and Bi4, as indicated.
The thin solid rectangle is the (2$\times$1) surface unit cell, and the dashed line is the mirror plane, which is the unique symmetry operation in this surface atomic structure.
}
\end{figure}

\begin{figure}[p]
\includegraphics[width=150mm]{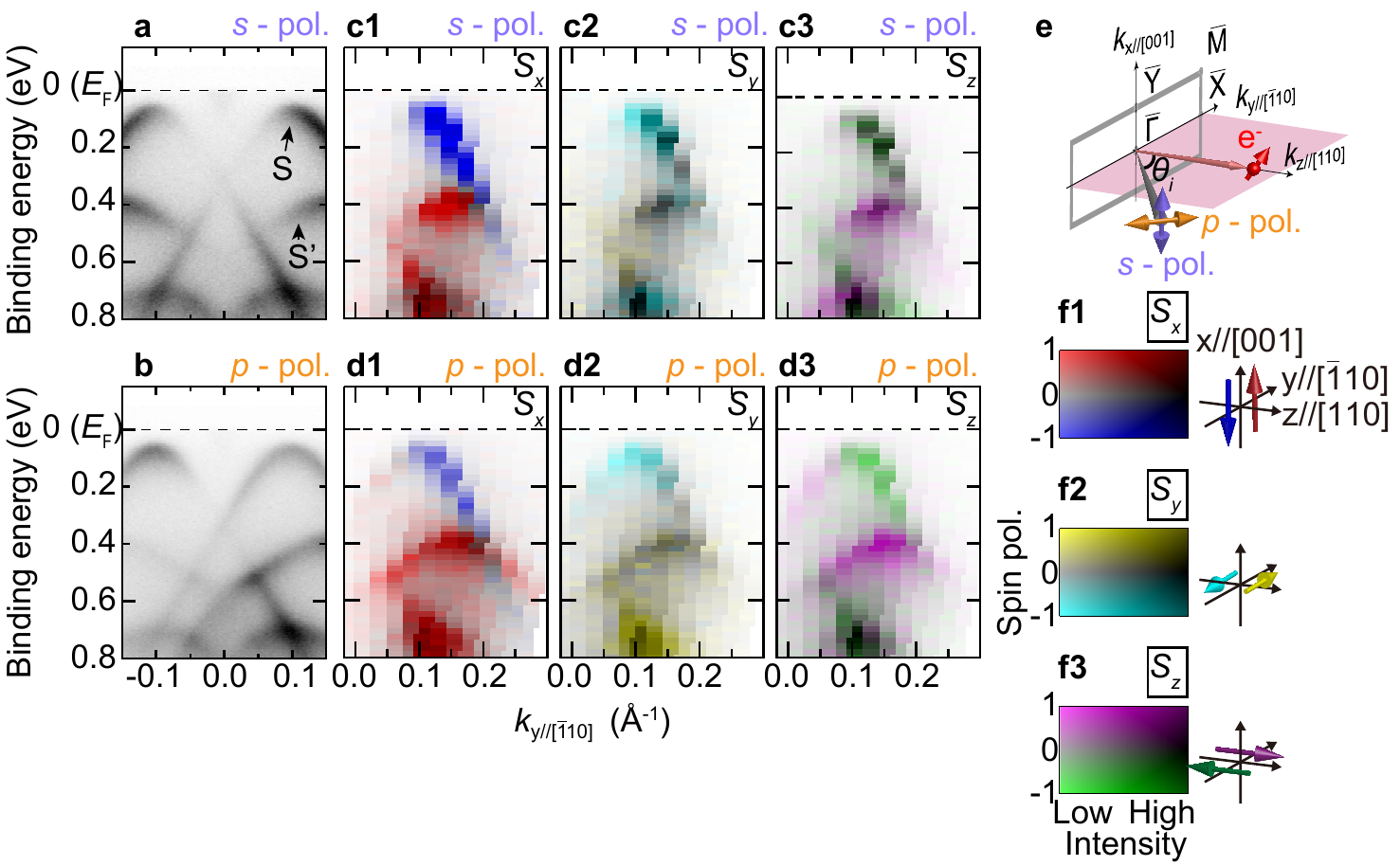}
\caption{\label{figure 2}
{\bf Spin-polarized surface band structures measured by SARPES.}
(a, b) ARPES and (c--d) SARPES 2D plots measured with (a, c1--c3) \textit{s}- and (b, d1--d3) \textit{p}-polarized photons along $\bar{\Gamma}$--$\bar{\rm X}$ (parallel to the Bi chains).
(a,b) Spin-integrated ARPES intensity maps. The photon incident angle ($\theta_i$) was $50^{\circ}$.
(c1--d3) SARPES maps polarized to $S_x$ (c1, d1), $S_y$ (c2, d2), and $S_z$ (c3, d3). The spin orientations are defined in (f). $\theta_i$ was $38^{\circ}$.
(e) Experimental geometry of the SARPES measurements and definitions of the coordinates.
The (2$\times$1) surface Brillouin zone and the common plane of the photon incidence and photoelectron detection are superposed simultaneously.
(f) Definitions of the spin directions of the photoelectrons.
The colors of the SARPES intensity plots are determined based on the photoelectron intensity and spin polarization, as indicated in these 2D color maps.
}
\end{figure}

\begin{figure}[p]
\includegraphics[width=150mm]{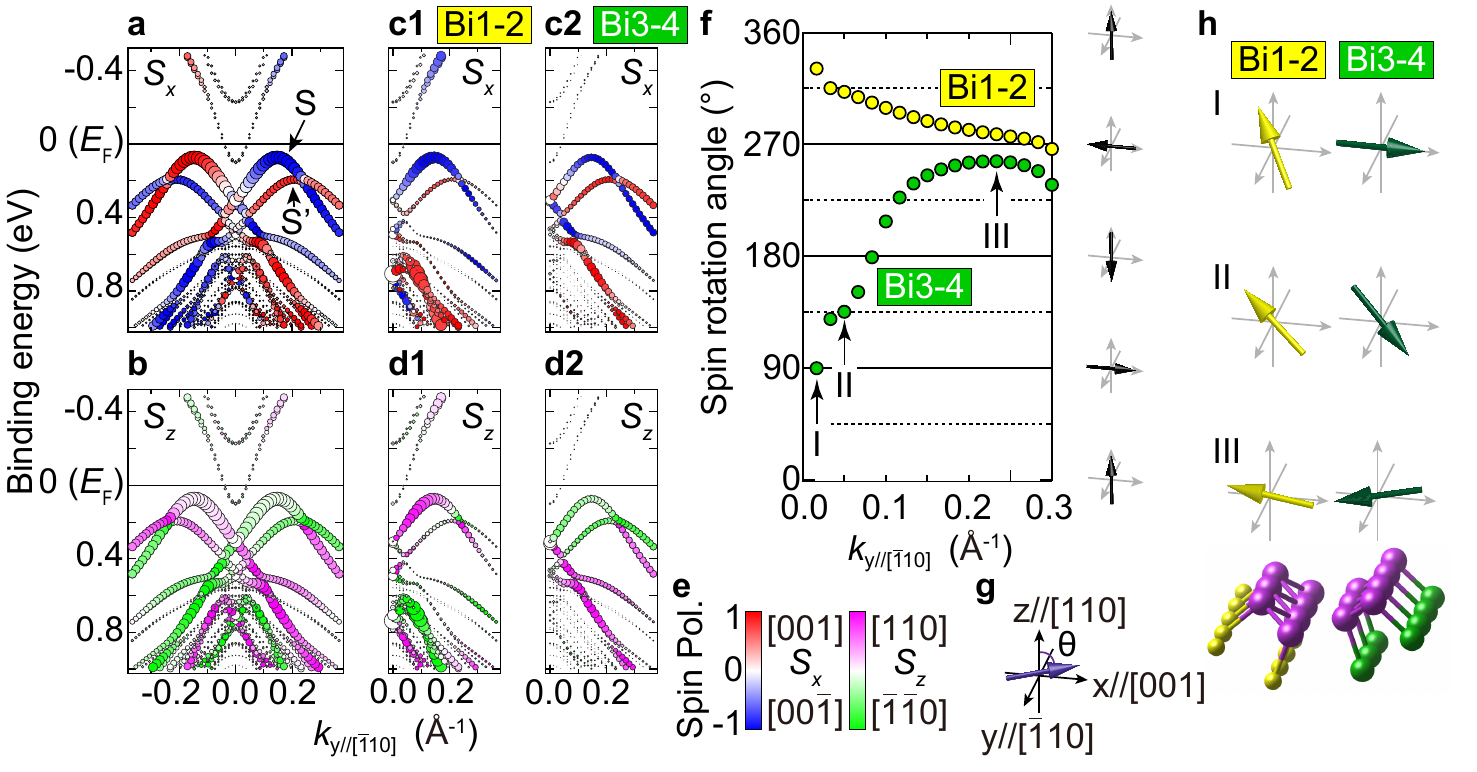}
\caption{\label{figure 3}
{\bf Calculated spin-polarized band structure and spin-texture in real space.}
(a--d2) Spin-polarized band dispersions obtained by DFT calculations.
The radii of the circles are proportional to the sum of the contributions from all surface Bi atoms.
Spin orientations are defined as in-plane ($S_x$, (a, c1--c2)) and out-of-plane ($S_z$, (b, d1--d2)) ones.
(c1--c2, d1--d2) Surface bands with the spin orientations of each non-equivalent surface Bi chain.
(e) Color scales represent spin polarizations: Blue--red (green--purple) color palette corresponds to the spins along the in-plane ($S_x$) (out-of-plane ($S_z$)) orientations.
(f) Wavenumber dependence of the spin orientations in each Bi chain.
(g) Definition of the spin rotation angle used in (f).
(h) Typical cases of the spin orientations depending on $k_{y//[\bar{1}10]}$ in each Bi chain. Roman numbers correspond to the $k_{y//[\bar{1}10]}$ position shown by arrows in (f).
}
\end{figure}

\begin{figure}[p]
\includegraphics[width=80mm]{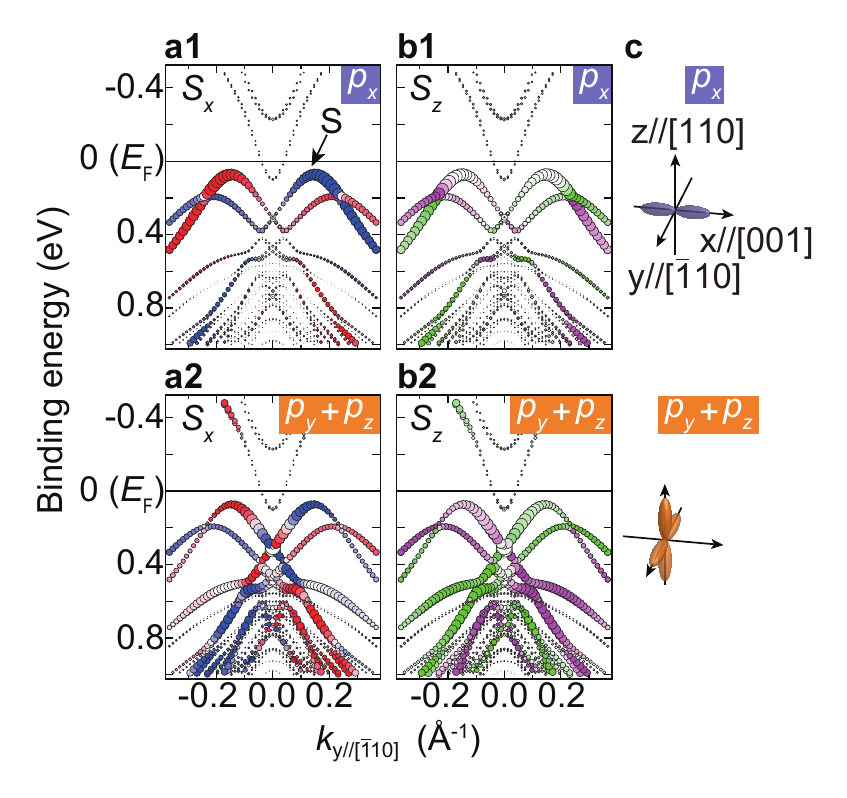}
\caption{\label{figure 4}
{\bf Spin polarizations of photoelectrons decomposed based on the symmetries.}
(a1--b2) Calculated band dispersions shown in a similar manner to Fig. 3 (a), but the radii of the circles are obtained by taking the Bi 6$p$ ($p_x$, $p_y$, and $p_z$) atomic orbitals, which have the odd (a1, b1) and even (a2, b2) symmetries with respect to the photon-incident plane in the SARPES experimental geometry.
The colors of the circles are from the corresponding Bi 6$p$ orbitals with $S_x$ and $S_z$ polarizations.
(c) Schematics of the Bi 6$p$ orbitals superposed on the coordinate axes.
}
\end{figure}

\begin{figure}[p]
\includegraphics[width=80mm]{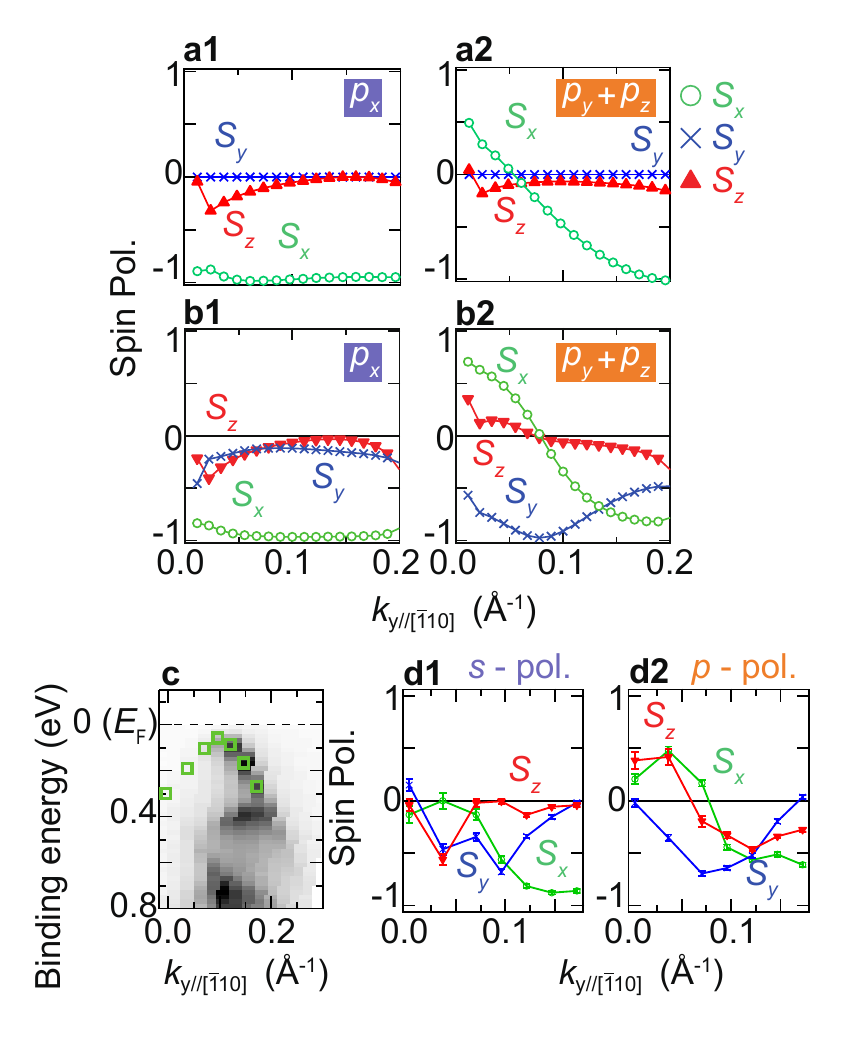}
\caption{\label{figure 5}
{\bf Spin polarizations of the photoelectrons from the computational and experimental analysis.}
(a1, a2) Three-dimensional spin polarizations of the initial state obtained from the DFT calculation with (a1) $p_x$ (odd) and (a2) $p_y$+ $p_z$ (even) Bi orbitals of the $S$ band.
(b1, b2) Photoelectron spin polarizations with (b1) $p_x$ and (b2) $p_y$+ $p_z$ Bi orbitals calculated numerically by considering the photoelectron interference of two non-equivalent Bi chains. 
The phase difference of the wave function between the Bi1-2 and Bi3-4 chains was set to 2$\pi$/3. See SM for details of the calculation \cite{SM}.
(c) ARPES band mapping with the \textit{s}-polarized photons.
Square rectangles represent the regions where the spin polarizations in (d1) and (d2) are obtained.
(d1, d2) Experimentally observed spin polarizations measured with (d1) \textit{s}- and (d2) \textit{p}-polarized photons.
}
\end{figure}

\renewcommand{\figurename}{Supplementary Figure}
\renewcommand{\thefigure}{S\arabic{figure}}
\setcounter{figure}{0}
\begin{figure}[p]
\includegraphics[width=80mm]{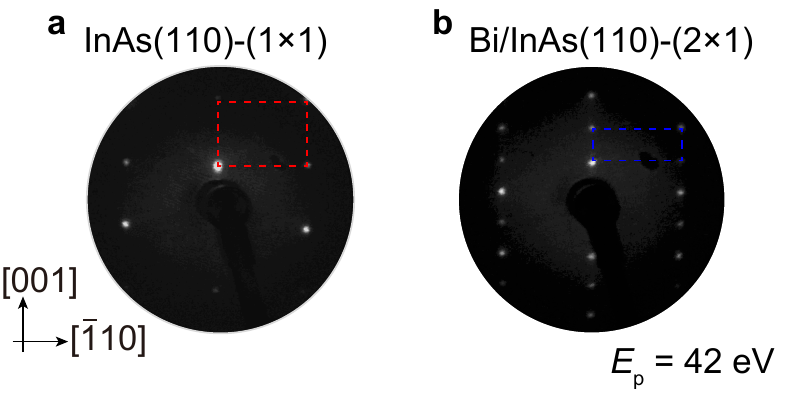}
\caption{\label{figure S1}
Low-energy electron diffraction patterns of (a) a clean InAs(110) substrate and (a) the Bi/InAs(110)-(2$\times$1) surface. 
Both images were taken at room temperature.
The red and blue dashed lines indicate (1$\times$1) and (2$\times$1) reciprocal lattices, respectively.
}
\end{figure}

\begin{figure}[p]
\includegraphics[width=150mm]{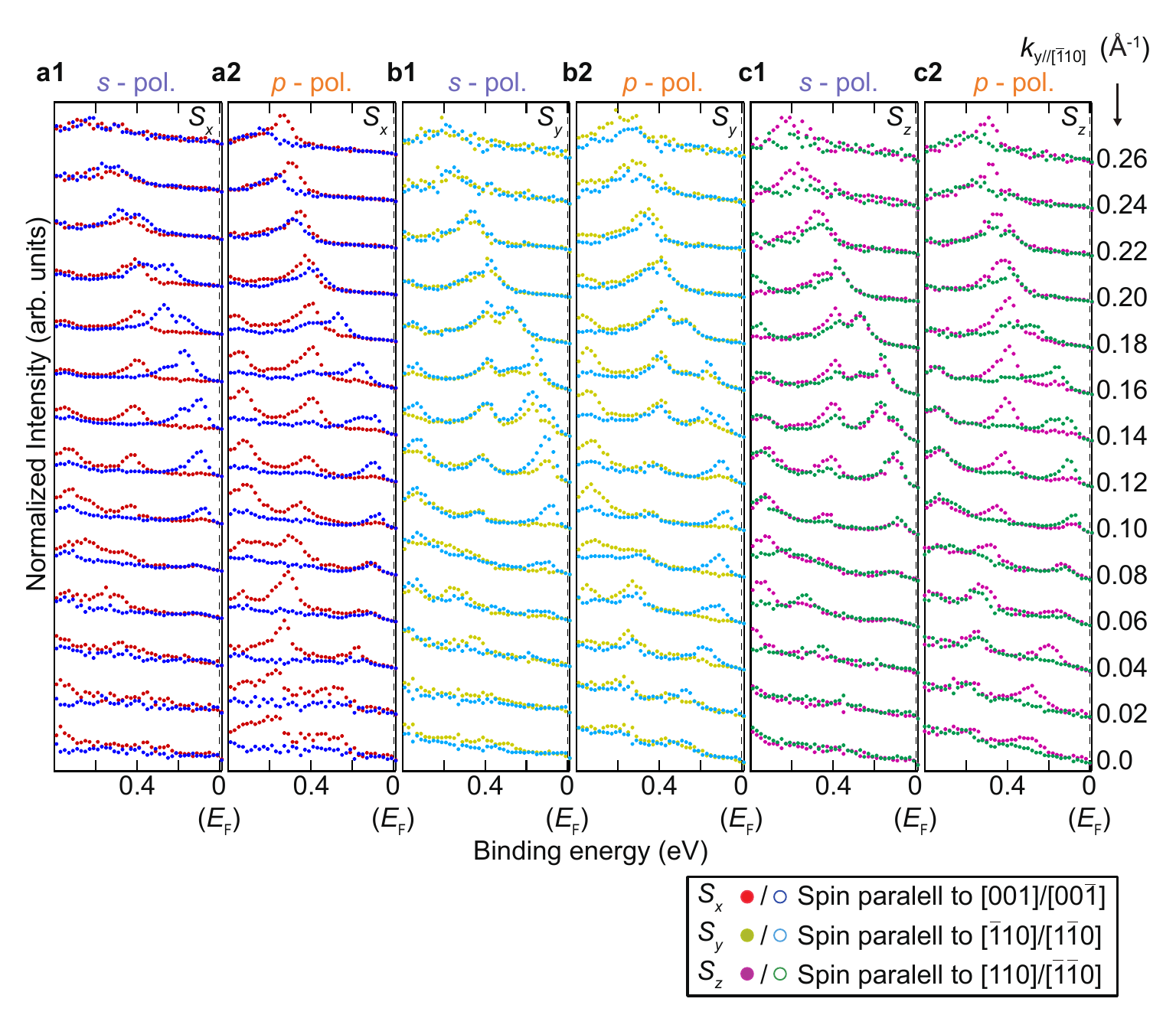}
\caption{\label{figure S2}
SARPES energy distribution curves (EDCs) with \textit{s}- (a1, b1, c1) and \textit{p}- (a2, b2, c2) polarized photons along $\bar{\Gamma}$--$\bar{\rm X}$.
The intensities are normalized by the maximum values of the spin integrated spectra at each $k_{y//[\bar{1}10]}$ in the measured energy range (binding energy = 0 $\sim$ 0.8 eV).
}
\end{figure}

\newpage

\section*{Supplementary Note 1. Numerical calculation of spin-polarized photoelectrons}
We describe the initial-state wavefunction as follows:
\begin{equation}
    \ket{i} = \phi_1\ket{\theta_1} + \phi_2\ket{\theta_2},
\end{equation}
where
\begin{equation}
  \ket{\theta_n} =  \left(
\begin{array}{c}
\rm{cos}\frac{\theta_n}{2} \\ 
\rm{sin}\frac{\theta_n}{2}

\end{array}
\right),
\end{equation}
represents the initial-state spin polarization of the \textit{n}th Bi chain (\textit{n} = 1, 2) pointing the angle $\theta_n$ in the \textit{xz} plane, and $\phi_n$ are the atomic wavefunctions from each Bi atom. 
Note that $\ket{0} (\ket{\pi})$ is the eigenspinor of the Pauli matrix $\sigma_z$ with an eigenvalue of 1 (-1).
Assuming the simple electric dipole transition to the nearly free-electron final state, the spinor field of the photoelectron $\chi$ is described by
\begin{equation}
    \ket{\chi_{\alpha}} = A_{1\alpha}\ket{\theta_1}+A_{2\alpha}\ket{\theta_2},
\end{equation}
where $\alpha$ = odd or even and $A_{n\alpha}$ is the complex matrix element from the atomic wavefunctions from the \textit{n}th Bi chain with $\alpha$ symmetry combined with the free-electron final state.

The spin expectation values of photoelectrons $P_{\beta}$ ($\beta = x, y, z$) are obtained with Pauli matrices $\sigma_{\beta}$ as follows:

\begin{equation}
    P_{\beta} = \frac{\bra{\chi_{\alpha}}\sigma_{\beta}\ket{\chi_{\alpha}}}{\braket{\chi_{\alpha}|\chi_{\alpha}}},
\end{equation}

\begin{eqnarray}
        \braket{\chi_{\alpha}|\chi_{\alpha}} &=& |A_{1\alpha}|^2 + |A_{2\alpha}|^2 + 2Re(A_{1\alpha}^*A_{2\alpha})\rm{cos}\frac{\theta_1-\theta_2}{2}, \\
       P_x\braket{\chi_{\alpha}|\chi_{\alpha}} &=& |A_{1\alpha}|^2\rm{sin}\theta_1 + |A_{2\alpha}|^2\rm{sin}\theta_2 + 2Re(A_{1\alpha}^*A_{2\alpha})\rm{sin}\frac{\theta_1+\theta_2}{2}, \\
      P_y\braket{\chi_{\alpha}|\chi_{\alpha}} &=& 2Im(A_{1\alpha}A_{2\alpha}^*)\rm{sin}\frac{\theta_1-\theta_2}{2}, \\
      P_z\braket{\chi_{\alpha}|\chi_{\alpha}} &=& |A_{1\alpha}|^2\rm{cos}\theta_1 + |A_{2\alpha}|^2\rm{cos}\theta_2 + 2Re(A_{1\alpha}^*A_{2\alpha})\rm{cos}\frac{\theta_1+\theta_2}{2}.
\end{eqnarray}

The first and second terms of $P_x$ and $P_z$ are from each initial state, and the third term originates from the interference between each Bi chain, namely, spin--orbital entanglement.
The same term also appears in $P_y$ where no initial state polarization exists.
Such photoelectron spin polarization from the interference during photoexcitation is similar to what has been reported for atomic wavefunctions in different layers [S1] and mixtures of even and odd orbitals [S2].

By setting the phase difference between $A_1$ and $A_2$ to be $\frac{2\pi}{3}$, we calculated the expected spin polarizations based on the ground states with the spin polarization orientations $\ket{\theta_n}$ and amplitudes of the atomic wavefunctions obtained by a DFT calculation.
As discussed in the main text, the obtained values shown in Figs. 5(d1) and (d2) in the main text agree well with the SARPES observation.

\section*{Supplementary Note 2. Low-energy electron diffraction patterns of InAs(110) and Bi/InAs(110)-(2$\times$1).}
Fig. S1(a) shows the low-energy electron diffraction (LEED) pattern of a InAs(110) substrate after Ar sputtering and annealing cycles.
Detailed conditions are described in the method section of the main text.
The Sharp and low-background pattern was observed, indicating a successfull surface cleaning. 
Fig. S1(b) shows the LEED pattern after Bi deposition and subsequent annealing as explained in the method section.
As shown by the blue dotted line, fractional order spots appeared, indicating the formation of the Bi/InAs(110)-(2$\times$1) surface.

\section*{Supplementary Note 3. Spin-resolved energy distribution curves taken along $\bar{\Gamma}$ -- $\bar{\rm X}$.}
Figures S2(a1, b1, c1) and S2(a1, b1, c1) show the polarization-dependent SARPES energy distribution curves (EDCs) by using \textit{s}- and \textit{p}-polarized photons, respectively.
The definitions of spin polarization ($S_x$, $S_y$, and $S_z$) are the same as those in Fig. 2(f) in the main text.
The intensities of SARPES EDCs are normalized by the maximum values of the spin integrated spectra (sum of the corresponding EDC pair for each $k_{y//[\bar{1}10]}$) in the measured energy range (binding energy = 0 $\sim$ 0.8 eV).

\subsection*{References of Supplementary Note}
\begin{itemize}
\item[S1.] Zhu, Z. -H. \textit{et al.} Photoelectron Spin-Polarization Control in the Topological Insulator Bi$_2$Se$_3$. \textit{Phys. Rev. Lett.} {\bf 112}, 076802 (2014).
\item[S2.] Yaji, K. \textit{et al.} Spin-dependent quantum interference in photoemission process from spin-orbit coupled states, \textit{Nature Commun.} {\bf 8} 14588 (2017).
\end{itemize}

\end{document}